\documentclass[doublecol]{epl2}  
\usepackage{epsfig} 
\title{Quantum plasticity and dislocation-induced supersolidity} 
\shorttitle{Quantum plasticity and dislocation-induced supersolidity} %Insert here a short version of the title if it exceeds 70 characters 
 
\author{Jean-Philippe Bouchaud \inst{1,2} and Giulio Biroli \inst{3}}  
\shortauthor{J.-P. Bouchaud, G. Biroli}  
  
\institute{                      
  \inst{1} Science \& Finance, Capital Fund Management, 6 Bd  
Haussmann, 75009 Paris, France.\\  
  \inst{2} Service de Physique de l'{\'E}tat Condens{\'e},  
Orme des Merisiers -- CEA Saclay, 91191 Gif sur Yvette Cedex, France.\\  
\inst{3} Service de Physique Th{\'e}orique, 
Orme des Merisiers -- CEA Saclay, 91191 Gif sur Yvette Cedex, France.  
}  

\abstract{We suggest that below a certain temperature $T_k$, the free energy for the creation of  
kinks-antikinks pairs in the dislocation network of solid He$^4$ becomes negative.   
The underlying physical mechanism is the related liberation of vacancies which initiate   
Feynman's permutation cycles in the bulk. Consequently, dislocations should wander and   
sweep an increasingly larger volume at low temperatures. This phenomenon should lead   
to a stiffening of the solid below $T_k$ and possibly to the appearance of a non zero superfluid fraction at a  
second temperature $T_c \leq T_k$.  }

\begin{document}  
\newcommand{\bq}{\ensuremath{{\bf q}}}  
\renewcommand{\cal}{\ensuremath{\mathcal}}  
\newcommand{\bqp}{\ensuremath{{\bf q'}}}  
\newcommand{\bbq}{\ensuremath{{\bf Q}}}   
\newcommand{\bp}{\ensuremath{{\bf p}}}  
\newcommand{\bpp}{\ensuremath{{\bf p'}}}  
\newcommand{\bk}{\ensuremath{{\bf k}}}  
\newcommand{\bx}{\ensuremath{{\bf x}}}  
\newcommand{\bxp}{\ensuremath{{\bf x'}}}  
\newcommand{\by}{\ensuremath{{\bf y}}}  
\newcommand{\byp}{\ensuremath{{\bf y'}}}  
\newcommand{\bxpp}{\ensuremath{{\bf x''}}}  
\newcommand{\rmd}{\ensuremath{{\rm d}}}  
\newcommand{\intk}{\ensuremath{{\int \frac{d^3\bk}{(2\pi)^3}}}}  
\newcommand{\intq}{\ensuremath{{\int \frac{d^3\bq}{(2\pi)^3}}}}  
\newcommand{\intqp}{\ensuremath{{\int \frac{d^3\bqp}{(2\pi)^3}}}}  
\newcommand{\intp}{\ensuremath{{\int \frac{d^3\bp}{(2\pi)^3}}}}  
\newcommand{\intpp}{\ensuremath{{\int \frac{d^3\bpp}{(2\pi)^3}}}}  
\newcommand{\intx}{\ensuremath{{\int d^3\bx}}}  
\newcommand{\intxp}{\ensuremath{{\int d^3\bx'}}}  
\newcommand{\intxpp}{\ensuremath{{\int d^3\bx''}}}  
\newcommand{\drho}{\ensuremath{{\delta\rho}}}  
\newcommand{\rhoh}{\ensuremath{{\widehat{\rho}}}}  
\newcommand{\fh}{\ensuremath{{\widehat{f}}}}  
\newcommand{\phih}{\ensuremath{{\widehat{\phi}}}}  
\newcommand{\thetah}{\ensuremath{{\widehat{\theta}}}}  
\newcommand{\etah}{\ensuremath{{\widehat{\eta}}}}  
\newcommand{\0}{\ensuremath{{(\bk,\omega)}}}  
\newcommand{\x}{\ensuremath{{(\bx,t)}}}  
\newcommand{\xp}{\ensuremath{{(\bx',t)}}}  
\newcommand{\xtp}{\ensuremath{{(\bx',t')}}}  
\newcommand{\xtpp}{\ensuremath{{(\bx'',t')}}}  
\newcommand{\xttpp}{\ensuremath{{(\bx'',t'')}}}  
\newcommand{\xtpn}{\ensuremath{{(\bx',-t')}}}  
\newcommand{\xtppn}{\ensuremath{{(\bx'',-t')}}}  
\newcommand{\xn}{\ensuremath{{(\bx,-t)}}}  
\newcommand{\xpn}{\ensuremath{{(\bx',-t)}}}  
\newcommand{\xppn}{\ensuremath{{(\bx',-t)}}}  
\newcommand{\xpp}{\ensuremath{{(\bx'',t)}}}  
\newcommand{\xxp}{\ensuremath{{(\bx,t;\bx',t')}}}  
\newcommand{\Crr}{\ensuremath{{C_{\rho\rho}}}}  
  
\newcommand{\Crf}{\ensuremath{{C_{\rho f}}}}  
\newcommand{\Crt}{\ensuremath{{C_{\rho\theta}}}}  
\newcommand{\Cff}{\ensuremath{{C_{ff}}}}  
\newcommand{\Cffh}{\ensuremath{{C_{f\fh}}}}  
\newcommand{\Ct}{\ensuremath{{\dot{C}}}}  
\newcommand{\Ctt}{\ensuremath{{\ddot{C}}}}  
\newcommand{\Crrp}{\ensuremath{{\dot{C}_{\rho\rho}}}}  
\newcommand{\Crfp}{\ensuremath{{\dot{C}_{\rho f}}}}  
\newcommand{\Crtp}{\ensuremath{{\dot{C}_{\rho\theta}}}}  
\newcommand{\Cffp}{\ensuremath{{\dot{C}_{ff}}}}  
\newcommand{\Crrpp}{\ensuremath{{\ddot{C}_{\rho\rho}}}}  
\newcommand{\thetab}{\ensuremath{{\overline{\theta}}}}  
\newcommand \be  {\begin{equation}}  
\newcommand \bea {\begin{eqnarray} \nonumber }  
\newcommand \ee  {\end{equation}}  
\newcommand \eea {\end{eqnarray}}  
\maketitle  
  
After a burst of theoretical activity in the early seventies \cite{Andreev,Leggett,Reatto}, the question of supersolidity is again the focus of intense attention since the discovery by Kim and Chan \cite{Chan1,Chan2} of a non classical moment of inertia in solid He$^4$ at low enough temperatures. The physical mechanism leading to superfluid flow in these crystals is still unclear; it is actually not even established that there is true
superflow. However, some consensus seems to prevail on several aspects of the problems \cite{Prokofiev}. From a theoretical point of view, there is now agreement that perfect He$^4$ crystals are {\it not} supersolids \cite{Ceperley,BPS1}, or at least have an extremely small superfluid fraction, much too small to account for experimental findings (see also   
\cite{Pomeau}). Two rather striking features have been emerged from experiments: i) the supersolid critical temperature and  the superfluid density are surprisingly large and ii) both depend sensitively on the detailed  preparation history of the crystal samples \cite{reppy1} and on the presence of minute fractions of He$^3$ impurities\cite{Chan1,Chan2}. The role of superfluid flow within grain boundaries \cite{Balibar}, possibly important in some experimental conditions, also seems to be moot since single He$^4$ crystals still appear to display significant supersolidity \cite{Chan2}. It now looks plausible that dislocations might play an important role. de Gennes \cite{PGG} has discussed some aspects of the quantum mobility of dislocations, and concluded that kink anti-kink bound states should suppress, rather than enhance, dislocation mobility. There has also been some recent work \cite{BPS2} suggesting that the {\it core} of dislocations could be superfluid, possibly leading to a (small) supersolid signal. A phenomenological theory of the role of dislocations can be found in \cite{Toner}. The aim of this note is to discuss some aspects of the physics of dislocations in a quantum crystal of bosons and suggest that below a certain temperature $T_d$, the free energy of kink-antikink pairs (in the climb direction) becomes negative, leading to an increasing wandering of dislocations that would form an entangled network similar to a polymer melt. This proliferation of kink-antikink pairs should lead both to an increased stiffness of the solid, and to kink-induced superfluid motion. The ideas put forth here are clearly highly speculative; we nevertheless hope to modestly contribute to the ongoing heated debate on the origin of supersolidity.    
  
Our approach follows Feynman's seminal paper on superfluidity \cite{Feynman}, and more recent work by Pollock and Ceperley \cite{CP} on the importance of permutation cycles to understand superfluidity. The starting point is the exact path-integral representation of the partition function $Z$ of a bosonic system of $N$ interacting particles as:  
\begin{eqnarray}  
Z &=& \frac{1}{N!} \sum_P \int \prod_i {\rm d}\vec z_i \int {\cal D}\vec x_i(t) \exp S(\{\vec x_i(t) \})\\  
S &=& \left[-\int_0^\beta {\rm d}t \left\{\sum_i \frac{m}{2\hbar^2}\left(\frac{\partial \vec x_i}{\partial t}\right)^2 + \sum_{i<j} V(\vec x_i-\vec x_j)\right\}\right]\nonumber  
\end{eqnarray}  
where the initial positions of the particles are $\vec z_i=\vec x_i(0)$'s and the final positions are constrained to be {\it any permutation} $P$ of the initial configuration. The difficulty is to correctly guess what set of configurations and ``trajectories'' dominate this very high dimensional integral.    
As emphasized by Feynman, although $t$ is the imaginary time, it helps intuition to think of a problem of $N$ classical particles moving in ``real'' time. At high temperature, $\beta=1/T$ is small, and the particles do not have much time to move around, so that the main contribution to $Z$ comes   
from the trivial identity permutation such that $x_i(0)=x_i(\beta)$, $\forall i$; the bosonic nature of the particles is irrelevant. When $\beta$ increases, more and more permutations can be explored. In a liquid,   
where motion is relatively easy, Feynman argues that at some temperature the average length of permutations diverges, leading to a superfluid phase transition. The precise connection with superfluidity   
was worked out by Ceperley and Pollock, who showed using linear response theory and periodic boundary conditions that the superfluid fraction $f_s$ can be exactly written as \cite{CP}:  
\be\label{fs}  
f_s = \frac{T}{T^*} \frac{N \langle \vec W^2 \rangle}{6a^2}; \qquad \vec W=N^{-1} \sum_i \int_0^\beta {\rm d}t\, \frac{\partial \vec x_i}{\partial t}  
\ee    
where the average is taken over all permutations with the weight defining $Z$, $a$ is the interatomic spacing (introduced for convenience) and $T^* \equiv \hbar^2/2ma^2$ is the typical kinetic energy of the particles ($a \approx 3$A  and $T^* \approx 1$K in solid He$^4$).   
 Any permutation can be decomposed into cycles. From the above formula, it is clear that only permutations corresponding to system spanning cycles wrapping around the torus   
can lead to a non zero winding number, $W$, and hence to a non zero superfluid fraction $f_s $. The weight of a given cycle involving the {\it simultaneous} jump of $n$ particles that all move a distance of order $a$ is:  
\be
p_n \sim \exp[-\frac{n \beta m}{2\hbar^2}\left(\frac{a}{\beta}\right)^2 - n \beta {\overline V}] = 
\exp\left[-n m^* \frac{T}{4T^*}\right],  
\ee
where ${\overline V}$ is a typical energy barrier encountered during the motion of each particle and $m^*=m \left(1+4 \overline V T^*/T^2 \right)$
is the effective mass. In reality the dependence of $m^*(T)$ on temperature is expected to be more complicated. In particular this expression is   
only valid at high enough temperatures. When $T < (T^* {\overline V})^{1/2}$, quantum tunneling sets in and leads to decrease of the effective value of ${\overline V}$. At zero temperature one expects a finite effective mass $m^*$, larger than the bare one $m$ because of interactions (using the previous notation this 
means that $\overline V$ vanishes quadratically with temperature as $T\rightarrow 0$).

Feynman's insight \cite{Feynman} was to realize that whereas $p_n$ decreases exponentially with $n$, the number of   
permutation cycles {\it increases} exponentially with $n$, as $z^n$ up to a power-law prefactor in $n$.  If the barrier ${\overline V}$ is small enough compared to $T$ (as in a liquid, where particles can  
move easily while avoiding each other), the entropy of these cycles dominates whenever $T < T_\lambda \simeq 4T^* \ln z -  {\overline V}$, favoring infinite cycles and hence superfluidity.   
In a solid, however, the energy of the intermediate state is much larger, and is in fact expected to grow {\it faster} than $n$ due to elastic deformations. Therefore, naively, only finite cycles are expected in a perfect solid, and the superfluid density remains zero.   
  
As has been recognized for a long time, the presence of vacancies can change this. Let us rephrase the argument of Andreev and Lifshitz \cite{Andreev}
at finite temperatures, which we will extend later to dislocation kinks. Vacancies in the initial configuration $\{z_i\}$ act as {\it initiators} of permutation cycles. A neighboring particle can easily hop to the empty site; once this is done (in imaginary time) the next particle can also easily hop, and so on (maybe slightly more collective moves are possible, too). For a path of length $n$, each particle must dash off to the neighboring empty site in ``time" $\sim 
\beta/n$, while paying an effective potential energy $\overline V$. The extra multiplicative weight associated to the path is now: 
\be  
p_n \sim \exp\left[- n \frac{\beta}{n} \frac{m}{2\hbar^2}\left(\frac{a}{\beta/n}\right)^2 - \beta {\overline V}\right] \equiv p_0 \exp\left[-n^2 \frac{T}{4T^*}\right].  
\ee   
In fact, the above estimate can be seen as a saddle point calculation where one decomposes the path in hops of duration $\tau_i$ such that  
$\sum_{i=1}^n \tau_i=\beta$, with fixed intermediate positions along the path. The saddle-point corrections add a factor $n C^n (T/T^*)^{3/2}/2\sqrt{\pi}$ to the above naive result, where $C$ is a numerical constant. The above calculation becomes unsuitable at low temperatures when transition 
between sites becomes instantonic transitions (i.e. localized events along the time axis). It leads to a slightly different expression for $p_n$ and allows one to recover the tight-binding model used by Andreev and Lifshitz (see Appendix). However, in both cases, $\ln p_n \sim - T^*/T G(n T/T^*)$, where $G(u)=G_0 + G_1 u+G_2 u^2$ for the above ``quasi-free'' model and $G(u)=u \ln (u/{\rm e}{\cal T})$ for the tight binding (instanton) model (see Appendix for the expression of the transmission coefficient $\cal T$). 
 
For the vacancy to return to its initial position at time $\beta$, the path must be a closed   
random walk on the crystal lattice. One can define the free-energy $F_v$ for the creation of a single vacancy {\it around a given site} as:
\footnote{The full free energy of a single vacancy should clearly include the obvious $-T \ln (V/a^3)$ contribution, corresponding to the
choice of the initial site on the permutation cycle.}
\be
F_v \approx E_v - T \ln \left[1 + \sum_{n=2}^\infty \Pi_n\right],\qquad \Pi_n\simeq  \frac{z^n p_n}{(2 \pi n)^{3/2}}  
\ee
where $E_v$ is the energy of a localized vacancy, when no permutation is allowed. The extra term   
${z^n}/{(2 \pi n)^{3/2}}$ accounts for the number of random walks returning at the origin after $n$ steps (for large $n$).  
Using again a saddle-point approximation for the sum, valid when $T \ll T^*$, we find that it is in both cases dominated by values of $n$ around  
$n^*=\mu T^*/T \gg 1$ with $\mu = 2\ln (zC)$ in the quasi-free model and $\mu = z{\cal T}$ in the instanton model. Noting that ${\overline V}(T \to 0)=0$, the limiting behaviours are easily found to be:   
\be\label{fv}  
F_v(T \ll T^*) \approx E_v - \zeta T^*; \qquad F_v(T \gg T^*) \approx E_v, 
\ee  
with $\zeta=\ln^2(zC)$ for the quasi-free model and $\zeta=z {\cal T}$ for the tight-binding model.
The leading correction when $T>0$ is found to be {\it positive}, equal to $+3T/2\, \ln T^*/T$ for both models, as it should be, since
finite temperatures inhibit permutation cycles. 
The free-energy for the creation a vacancy $F_v$ is therefore lowered at low temperatures. The shape of $F_v-E_v$ as a function of $T/T^*$,  
for both models and for different choice of parameters is shown in Fig. 1.\footnote{In a fermionic environment, the free-energy reduction is much weaker, because different permutations interfere destructively. As a rough approximation one may keep only self-retracing paths, which leads to a 
reduction of $\mu$ and $\zeta$ in the above formulas.} Depending on the value of these parameters, one may find a temperature $T_v$ at which 
$F_v$ vanishes. When $F_v >0$, the equilibrium density $\phi_v$ of vacancies is given obtained by minimizing the total free energy (per unit volume):
\begin{equation}
{\cal F} =  T \phi_v \left[ \ln \left(\frac{\phi_v a^3}{{\rm e}}\right)-1\right] + \phi_v F_v +O(\phi_v^2) \quad.
\end{equation}
This leads to 
\be
\phi_v^* = \frac{\exp(-F_v/T)}{a^3}, 
\ee
which is vanishingly small at low temperatures. If on the other hand $F_v$ becomes negative, one expects a finite density of vacancies in the system
even when $T \to 0$. This density is now controlled by the vacancy-vacancy interaction \cite{Andreev}, which adds a term $A \phi_v^2/2$ to the
above expression for ${\cal F}$. When $T_v \ll T^*$, the vacancy density first decreases as the temperature is reduced, but then 
sharply increases as the temperature approches $T_v$. At low temperatures, the density is given by $\phi_v^* \approx |F_v|/A$. When $A$
is small, the proliferation of vacancies might actually lead to a true phase transition to a low density solid phase.
The case of kinks discussed below will closely follow the above argument. Note that the decrease of $\phi_v$ until $T_v$ holds  
only when $F_v$ decreases faster than linearly with temperature, which is indeed what we finds in our approximate treatments, see Fig. 1. 
Even when this is not the case, $T_v$ is the characteristic temperature determining the point where vacancy-vacancy interaction 
becomes crucial in order to limit the density of vacancies (which would otherwise proliferate). 
\begin{figure}  
%\onefigure{edge.ps}  
\psfig{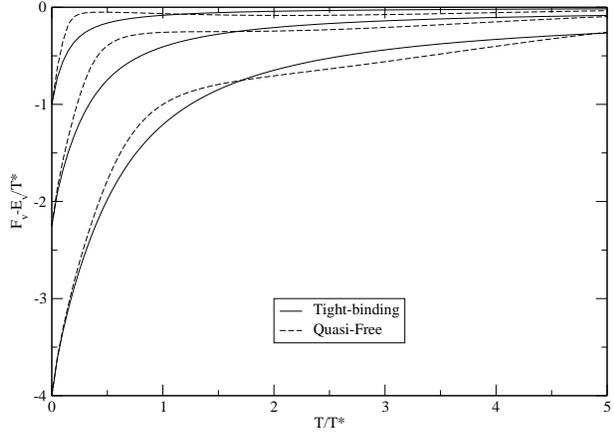}   
\caption{Evolution with temperature of the excess free-energy $F_v-E_v$ for the creation of a single vacancy, lowered by the surrounding permutations   
of the bosonic solid, for both models: quasi-free (dashed lines) and tight binding (plain lines). From top to bottom, $\zeta=1^2,1.5^2,2^2$ and ${\overline V}=0$ for the quasi-free model (this is only reasonable at low enough $T/T^*$).  Axis are in reduced units: $T/T^*$ and $F/T^*$.}  
\end{figure}  
Now, for superfluidity to set in, the permutation paths should not end where they where initiated since in this case the winding number $W$ in Eq. (\ref{fs}) above is zero. Paths should rather end on the  
site where a different vacancy started off. If the end-to-end distance of the path is $R$, the corresponding weight is modified to:  
\be  
\Pi_n(R) =  \Pi_n \exp\left[- \frac{R^2}{2 n a^2}\right].  
\ee  
It is clear that these paths are exponentially suppressed when $R^2 \gg n^* a^2$. If one wants to see the appearance of a permutation cycle winding around the system and contributing to  
$W$, the distance between vacancies should be such that the suppression factor does not preclude the existence of a percolating path of inter-vacancy hops, which happens below a certain   
temperature $T_c$ given by:  
\be\label{Tc}  
a\sqrt{n^*} \sim a\phi_v^{*-1/3} \longrightarrow T_c \sim  \frac{\mu}{2} \phi_v^{*2/3} T^*.  
\ee  
The superfluid transition temperature $T_c$ is of the same order of magnitude as the temperature at which a dilute gas of particles of mass $m$ would Bose condense.   
Below $T_c$, the superfluid fraction can be estimated using the Ceperley-Pollock formula. The probability that a given atom belongs to a winding chain is $\phi_v^* n^*$, each of which contributes to  
$\langle W^2 \rangle$ by an amount $a^2/N$, leading to:  
\be  
f_s \approx \frac{T}{T^*} \phi_v^*  \frac{\mu T^*}{2T} \sim \phi_v^*;  
\ee  
i.e. the superfluid density is of the order of the vacancy density, which is itself 
temperature dependent. Due to the large positive value of $F_v(T=0)$ in solid He$^4$ (around $13 K$, see \cite{BPS1}), the density of vacancies is extremely small and vanishes at zero temperature \cite{BPS3}. As a consequence, 
the condition (\ref{Tc}) is never met. No supersolid transition induced by vacancy delocalization is expected and 
this scenario cannot account for experimental results.   
  
On the other hand, it is most probable that even single He$^4$ crystal contains quenched-in dislocations; estimates vary in the range $10^6-10^{10}$ cm$^{-2}$. This means that the typical distance $D$  
between two dislocations is in the range $300 \leq D/a \leq 3\, 10^{4}$, and the probability $\phi$ for an atom to be part of a dislocation of the order $\phi \sim 10^{-9}-10^{-5}$, much smaller than the superfluid fraction   
$2 \, 10^{-3}$ reported for isotropically pure single crystals (1 ppb of He$^3$ impurities). An important message conveyed by the work of Boninsegni et al. \cite{BPS2} is that dislocation cores can be considered as liquid-like, and  
permutation cycles are favored {\it along} the dislocation lines at low temperatures. Indeed, even classically self-diffusion is enhanced in dislocation cores \cite{tucker}.  
Because these lines are one dimensional, however, superfluid coherence is only maintained up to length $\ell$ such that $\ell \sim a T^*/T$ (see, e.g. \cite{Toner}).   
  
%\begin{figure}  
%\onefigure{epl-template.eps}  
%\caption{Figure caption.}  
%\label{fig.1}  
%\end{figure}  
  
\begin{figure}  
%\onefigure{edge.ps}  
\psfig{file=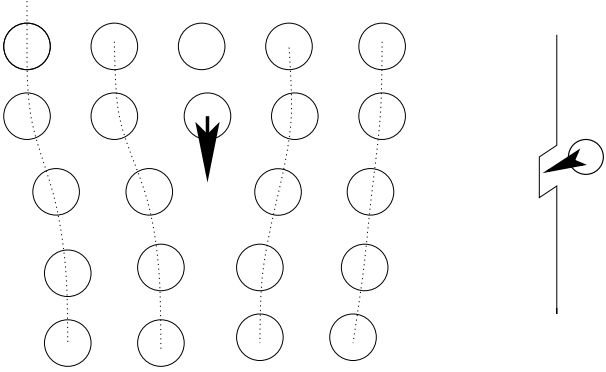,width=8cm}   
\caption{Left: cartoon of an edge dislocation in the plane perpendicular to the dislocation direction. The creation of a kink-antikink in the climb direction (or jog) corresponds to liberating a vacancy, as indicated by the arrow, which can then initiate a permutation cycle in the bulk. Right: side view   
of the same situation.}  
\end{figure}

Now, consider an edge dislocation with a kink-antikink corresponding to one (or several) atom(s) moving in the climb direction (see Fig. 2) \footnote{These objects are in fact called jogs in the dislocation literature \cite{Hirth} but we will still call them kinks.}.  In effect, this creates a vacancy (or a string of vacancies) which can now act as  
the starting point of permutation cycles in the direction {\it transverse} to the dislocation. 
The creation energy of a kink-antikink pair, $E_k$, is expected to be appreciably lower than that of a vacancy $E_v$, 
since the crystal is already strongly deformed around the dislocation.  When temperature goes down, the creation free-energy of the kink-antikink pair 
is lowered by the permutation cycles initiated around it, and is given by an expression similar to Eq. (\ref{fv}):  
\be\label{fk}  
F_k(T \ll T^*) \approx E_k + {\overline B}(T) - \zeta T^*,  
\ee  
where ${\overline B}(T)$ is the effective barrier to unbind the vacancy from the dislocation core, averaged over the relevant vacancy paths and corrected
by quantum tunnelling effects. At high temperature it is equal to the binding energy, $E_B$, between the vacancy and the dislocation core, but is
reduced by quantum fluctuations at low temperatures. This effective barrier in fact depends on the typical extension of   
the paths $R^* \sim a\sqrt{n^*}$ and on the orientation of these paths, since the deformation field around the dislocation contains both 
compressed and expanded regions \cite{Toner}. 
  
Our scenario relies on the assumption that $E_k, {\overline B}$ are small enough such that $F_k(T)$ becomes negative below a temperature $T_k$ (see Fig 1). Both $E_k$ and $\overline B$ should indeed not to be larger than a few Kelvins, at least when pressure is not too large.   
Taking for example $\zeta=4$, $E_k=3T^*$, and ${\overline B}(T \to 0) \approx 0$ leads to $T_k \sim 0.2 T^*$ or 200 mK (see Fig. 1), 
with a rather modest extension of permutation cycles $R^*(T_k) \sim 3 a$: the released vacancies do not travel very far away from the dislocation cores. 
If the free energy for the creation of kinks indeed become negative below $T_k$, one should see a proliferation of kinks and anti-kinks in the configurations that contribute most to the partition function $Z$ of the solid\footnote{There is a remark in   
de Gennes' paper \cite{PGG} where this possibility is mentioned, which he tentatively associates with melting of the solid.}. This would lead to a substantial lengthening of the already present dislocations, which start wandering around in the solid, carrying an $O(1)$ density of kinks. Bose statistics of the surrounding atoms favors the extension of the curvilinear length of dislocations.   
%which become truly delocalized, quantum objects.   
The nature of the resulting quantum dislocation ``soup'', i.e. how much dislocations can wander around into the solid and 
how large are the regions sweeped by dislocations, is a very complicated problem. The difficulty is to give a correct 
quantitative treatement of quantum dislocations. We suggest that the the limiting factor to a complete delocalization 
of dislocations is due to the elastic energy. Consider for example   
a single dislocation. From the reasoning above, it should carry, below $T_k$ an $O(1)$ density of kinks-antikinks pairs.   
Large fluctuations can only be achieved if kinks and antikinks unpair. However, in this case one has to consider on top of the free energy (\ref{fk}) 
the elastic interaction energy between kinks and antikinks which disfavors the accumulation of kinks and the corresponding wandering of the   
dislocation on large length scales. Furthermore dislocations form a network inside the solid. Thus, one also has to consider the elastic   
energy due to dislocation-dislocation interactions which, again, limits the density of dislocations. Other (shorter-range) 
interactions between kinks and antikinks could also play an important role.

We propose here an admittedly very naive analysis of this difficult problem. Calling $\phi_d$ the total density of sites   
sweeped by dislocations (per unit volume now), a simple mean-field argument allows one to write the free-energy per unit volume   
of the dislocation network as:   
\be\label{freeenerg}  
f \approx F_k  \phi_d  + \frac{1}{2} Y \phi_d^2  
\ee  
where $Y$ is a typical elastic interaction scale, formed with the shear modulus $G$ and the atomic volume $a^3$.   
Since $Ga^3 \approx 100$K, a rough order of magnitude is $Y \sim 50$ K, to within a factor 2.  The equilibrium density of dislocations is therefore $\phi_d^* \sim  -F_k/Y$. Below $T_k$ there is thus an extra contribution $df$ to the total free energy coming from the dislocation soup and the  corresponding proliferation of permutation cycles. Using the above naive estimate, the extra free energy contribution reads, close to $T_k$: $df \sim -(T_k-T)^2$, leading to a small extra specific heat contribution below $T_k$. Whether $T_k$ corresponds to a true phase transition where the dislocations network changes nature and forms a kind of  
``quantum dislocation soup'' phase is an open problem. Well below $T_k$, a reasonable estimate is $F_k  \sim -50$ mK, leading to $\phi_d^* \sim 10^{-3}$, a significant increase from the bare estimate $10^{-9}-10^{-5}$ given above, which assumes that dislocations are essentially straight lines. However, the density of the system, or the Debye-Waller factor, should only change by very small, unmeasurable amounts at $T_k$, compatible with experimental findings \cite{DW}. Since the density of liquid and solid only differ by $10 \%$, we expect the change of density at $T_k$ to be $\sim 10^{-4}$ at most. But the delocalization of dislocations should affect the elastic property of the solid. From the above argument, the density and fluctuations of the ``polymer melt'' are governed by the repulsive (elastic) interactions between otherwise proliferating and expanding dislocations.   
When the solid is deformed, the melt must adapt and deform as well. Since the initial state minimizes the free energy, Eq. (\ref{freeenerg}), any shear deformation can only increase the free energy to quadratic order, leading to an increased shear (and bulk) modulus. This might explain the recent results from the Beamish group \cite{Beamish}, the anomaly in the sound velocity reported some years ago by Goodkind et al \cite{Goodkind} and the elastic resonance frequencies reported in \cite{Mukharski}. The elastic region should however significantly narrow down below $T_k$, giving way to enhanced plasticity effects. We note at this stage that the sudden appearance of a dislocation melt could lead to partial decoupling of the solid in a   
oscillating pendulum experiment \cite{Nussinov}, although more careful calculations are needed to see if this is enough to explain the observed NCRI fraction.  
  
Coming back to a possible supersolidity transition, we follow Boninsegni et al. \cite{BPS2} who show that superfluid order establishes along the (coiled) dislocations. Using the kink mechanism above superfluid order can also explore a sausage of radius $R \sim a \sqrt{n^*}$ around the sites sweeped by dislocations. This can be thought of as a kink-induced proximity effect. Superfluidity propagates from dislocation to dislocation over the whole system only if    
another dislocation is typically present within a sausage of coherence length $\ell \simeq a T^*/T$ (a kink would move very easily along the dislocation to match the incoming permutation path). The condition for macroscopic superfluidity therefore reads:  
\be  
\phi_d^* \times (\pi R^2 \ell) > 1 \to T < T_c = T^* \sqrt{\mu \pi  \phi_d^*/2} \sim 100 \mbox{mK},  
\ee     
which would give the right order of magnitude for the appearance of supersolidity in experiments. Using the Pollock-Ceperley formula, the superfluid density is then given by the ``saugage'' fraction (each dislocation point can be the source of a permutation cycle up to a fast translation of a kink) :  
\be  
f_s =  \frac{T}{T^*} \phi_d^*  \frac{\mu T^*}{6T} \sim  3 \, 10^{-4},  
\ee  
which is compatible with recent data on single crystals with 1ppb  He$^3$ impurities. Clearly, the above numbers are only intended to be rough estimates, maybe more significant is our unusual, 2d like scaling prediction $T_c \sim T^* \sqrt{\phi_d^*}$.   
  
An interesting consequence of our scenario is the role of He$^3$ impurities. Calling $\varphi$ the fraction of these impurities and assuming that they mostly gather   
within dislocation cores at low temperatures, one finds that the   
typical distance between He$^3$ impurities along dislocations is $d = a \phi_d^*/\varphi$. In Feynman's picture, it is clear that He$^3$ are detrimental to superfluidity since any permutation cycle   
involving an He$^3$ atom will not contribute to the partition function anymore. One expects that as soon as $d \sim \ell$,  the concentration of  He$^3$ becomes noticeable and affects the superfluid  
density.  When $T \sim 10$ mK, our estimate above leads to $\ell/a \sim 100$ and hence $\varphi=10$ ppm. On the other hand, He$^3$ impurities might also change the density of quenched-in dislocations  
and reduce the bare energy of a kink-antikink $E_k$, so that the overall influence of He$^3$ on supersolidity could be rather complex. 
 
What about {\it solid} He$^3$? The delocalisation of vacancies at low temperatures, and therefore the lowering of the kink-antikink creation  
energy, may also occurs in this case. However, the effect is much weaker for fermions and $T_k$ is expected to be much smaller than in He$^4$ (see footnote 2). Kinks should proliferate more easily in spin polarised than in unpolarised solid He$^3$. Of course, there is no superfluidity in this case (except if vacancies pair up), but there could still be observable elastic anomalies in solid He$^3$ at low enough temperature and pressure (perhaps also accompanied by a small magnetic susceptibility anomaly due to the formation of spin-polarons around vacancies \cite{Lhuillier}). More experiments on 
the elastic properties of solid He$^3$ would be welcome.
  
Let us summarize the main features of our scenario. Around a temperature $T_k$, kinks proliferate and the dislocations form a ``quantum dislocation soup'' 
analoguous to an entangled polymer melt. Whether the dislocations network undergoes a true phase transition is at this stage an open problem, but one expects 
a (small) specific heat anomaly and a change the elastic properties of the system around $T_k$. This could explain the shear modulus anomaly recently reported by Day and Beamish \cite{Beamish} and other elastic anomalies, perhaps even a partial decoupling of the solid in a   
oscillating pendulum experiment. If $T_c < T_k$, one should see a region with modified shear (and bulk) modulus but only a small superfluid density before supersolidity really sets in at $T_c$, when permutation cycles hook up different dislocations.  Because our mechanism is a hybrid between longitudinal (along dislocation lines) and transverse permutation cycles, small concentration of He$^3$ impurities may significantly reduce the superfluid density. If the kink-energy can be increased   
substantially by increasing the pressure, one should see a complete suppression of both elastic anomalies and supersolidity when $E_k(P) > \zeta(P) T^*$. On the other hand, one should also take into account how the initial concentration of dislocations changes upon increasing the pressure.   
The mechanism for supersolidity proposed in this work may also be relevant in more general cases. Indeed, supersolidity has been obtained   
not only for single He$^4$ crystals but also for He$^4$ in porous Vycor \cite{Chan1} and after rapid quenches \cite{reppy2}. In these  
two latter cases, the underlying solid matrix could be so full of defects that taking the crystal as the   
reference state and considering its defects such as dislocations may not be relevant. A more sensible reference state might be an amorphous solid \cite{superglass}. Nevertheless, supersolidity could also arise in this case because of the liberation of vacancies from ``soft'' preferred regions following a mechanism similar to the one we proposed in this work. Finally, we suggest that the repulsive interaction between superfluid vortices and coiled dislocations could lead to interesting physical effects; we expect shear cycles below $T_c$ to be hysteretic and affect the supersolid properties.

\acknowledgments  
We thank S. Balibar for many discussions and encouragements. Comments by F. Caupin, P. Goldbart, Y. Mukharsky,   
E. Varoquaux, M. Wyart and F. Zamponi were also most useful. This paper is dedicated to the memory of PG de Gennes, from whom we learnt so much and with whom we would have liked to discuss the ideas presented here, which were triggered by his own work.  
  
\section{Appendix}

In the following we present a derivation of the excess free energy 
for a vacancy in a bosonic crystal, $F_v$, which makes clear the connection with
Andreev and Lifshitz \cite{Andreev}. 
We consider the contribution to the path integral given by paths in which 
a vacancy is created in the origin just after $t=0$, it wanders around and come back at the origin at 
$t \approx \beta$. We estimate the additional multiplicative weight induced by this process 
compared to the one in which particles oscillate around crystalline positions. Close to $t=0$ 
a vacancy-interstitial pair is formed. The potential energy of this configuration is higher than 
the initial crystalline one. Once created and at 
low temperature, vacancies allow the system to gain
kinetic energy by moving through the lattice. 
In order to estimate this effect we note that the crystal with the 
vacancy-interstitial pair is a local minimum of the potential energy. Moving the vacancy on other
lattice sites one obtains new local minima with the same energy 
(at least if the vacancy is far enough from the interstitial which will happen most of the time at large $\beta$).
As a consequence the extra weight due to paths in which the vacancy starts at $t\approx 0$ and comes 
back at the same position at $t\approx \beta$ can be estimated summing over instantons, where
each instanton corresponds to the motion of the vacancy to a nearest neighbor lattice site (we neglect
for simplicity all other jumps). Each step leads to a factor \cite{Zinn-Justin} $K=T^* g \exp (-S_0)$,
where $S_0$ is the value of the action for a single instanton and $g$ is 
the contribution due to the Gaussian integration around the saddle point. We can write $K=T^* {\cal T}$,
where ${\cal T}$ is the transmission coefficient of the barrier. For large $\beta$ one has to sum a dilute gas of instantons. 
The integration over the instanton positions in time gives, for $n$ instantons, $(\beta K)^n/n!$ \cite{Zinn-Justin}. 
Thus, the weight of a path of $n$ steps on the lattice can be written (for large $n$): 
\begin{equation}  
p_n=\frac{C}{\sqrt{2\pi n}}\exp\left[-\frac{T^*G(nT/T^*)}{T}\right], \quad G(u)= u \ln (u/{\rm e}{\cal T})
\end{equation}
where the constant factor $C$ takes into account the extra contribution due to the creation and annihilation
of the interstitial-vacancy pair at $t \approx 0,\beta$.
The computation of $F_v$ via the sum over closed paths of length $n$ can be performed by decomposing over paths   
consisting of $n_x$ forward and $n_x$ backward steps in the $x$ direction (and similarly for the other two directions). 
The number of closed paths of $n$ steps with $n_x,n_y,n_z$ forward steps in the $x,y,z$ directions, is equal to 
$n!/(n_x!^2n_y!^2n_z!^2)\delta_{n,2n_x+2n_y+2n_z}$ (for simplicity we focus on a cubic lattice). 
Introducing into the sum the identity 
$f(n)^{2}=\int_0^{2\pi} d\theta/(2\pi)\sum_{n'=0}^{\infty}e^{-i\theta(n-n')}f(n)f(n')$ and summing over all
$n$s variable one finds the familiar result: 
\begin{equation}  
F_v = E_v - 3T \ln \left(\int_0^{2\pi} \frac{d\theta}{2\pi} \exp\left(2\beta K \cos(\theta) \right)\right)
\end{equation}
Up to proportionality constants, this is exactly what one would have obtained for a particle (vacancy)
in a tight binding model at temperature $T$. This computation allows one to show how vacancies can lead to 
quasi-particle excitations and bridges the gap between our approach and the one of Andreev and Lifshitz \cite{Andreev}
who analyzed the $T=0$ case starting directly from a tight binding model. Note that $F_v(T=0)=E_v-6T^* {\cal T}$.

\end{document}